\documentclass[12pt]{iopart}
\usepackage{graphicx}
\usepackage{iopams}
\usepackage{setstack}

\newcommand{\vecd}[1]{\bi{#1}}
\newcommand{\matr}[1]{\sf{#1}}

\begin{document}

\title{Hydrodynamic attraction and repulsion between asymmetric rotors}

\author{Steffen Schreiber$^1$, Thomas Fischer$^2$ and Walter Zimmermann$^1$}

\address{$^1$Theoretische Physik I, Universit\"at Bayreuth, D-95440 Bayreuth, Germany\\
          $^2$Experimentalphysik V, Universit\"at Bayreuth, D-95440 Bayreuth, Germany
}
\ead{walter.zimmermann@uni-bayreuth.de}
\date{\today}

\begin{abstract}
At low Reynolds numbers, the hydrodynamic interaction between dumbbells driven by an external rotating field can be attractive or repulsive.
Dumbbells of dissimilar asymmetric shape or different coupling to the external field undergo conformational rearrangements that break the time reversal symmetry.
The parameter ranges leading to attraction or repulsion are explored numerically. The results of our simulations suggest that rotating fields
 may be a useful avenue for the assembly, disassembly, and sorting of particles of different shape as well as for the study of collective micro-swimmers.
\end{abstract}

\section{Introduction}\label{sec: intro}

Hydrodynamic interactions in suspensions of active 
or externally driven nanoparticles or microorganisms are 
involved in collective, biological, chemical and technical 
motion \cite{GITaylor:1951.1,Brenner:1981,Berg:1973.1,Purcell:1977.1,Kessler:1992.1,Stone:2001.1,Grzybowski:2004.1,Ishikawa:2006.1,Lauga:2009.1}
of fascinating complexity.
The propulsion of micro-particles and microorganisms differs substantially from the motion
of macroscopic objects like fishes or birds due to the small scale motion at low Reynolds numbers \cite{Brenner:1981}.
As a consequence swimming macro- and microorganisms transform chemical
into translational or rotational energy in different ways.
Considerable progress has been made
in understanding the propulsion of microorganisms
by studying artificial active swimmers
 \cite{Lauga:2009.1,Stone:2003.1,Dreyfus:2005.2} or active
rotors  \cite{Dreyfus:2005.1}.
Self-organization phenomena of  assemblies of micro-swimmers can teach us
  how hydrodynamic  interactions synchronize the motion of swimmers or cilia
 The importance of hydrodynamic
interactions (HI)  at small Reynolds numbers has been
emphasized by showing that they may cause a lift force
during the motion of vesicles close
to a wall \cite{Misbah:99.1,Seifert:99.1,Viallat:2002.1}. They also lead to cross-streamline migration of droplets,  vesicles
and deformable bead-spring models in Poiseuille flow \cite{Leal:1980.1,Kaoui:2008.1,Arend:2008.1}.
Additionally, the interplay between Brownian motion of small
particles  and hydrodynamic interactions
 can cause surprising anti-correlations and cross-correlations
\cite{Quake:99.1,Ziehl:2009.1}.
Brownian dynamics of hydrodynamically interacting  polymers
may cause
elastic turbulence that is used for mixing in microfluidic devices  \cite{Groisman:2000.1,Groisman:2001.1}.

Macroscopic rotors experience the attractive Magnus effect that can easily be understood using the concept of dynamic pressure.
The attraction or repulsion between rotors in the limit
of Stokes flow is less obvious.
Magnetically driven rotors can attract each other due to their magnetic
interaction \cite{Stone:2001.1,Grzybowski:2000.1,TFischer:2007.1}. They form
dynamically induced self-organized structures
depending on the particles' contours \cite{Grzybowski:2004.1}.
Magnetic dipole-hexapol interactions
may be used for separating
rotors with respect to their shape \cite{TFischer:2009.1}.
Rotors also play an important role for
bacterial motion such as  the propulsion of
Escherichia coli \cite{Berg:1973.1} with its rotating flagella.
The dynamics of single flagella has recently
been modeled in numerical calculations \cite{Powers:2004.1}.
There hydrodynamic interactions are the source of the propulsion the
flagella create.
One fundamental problem for active and driven low Reynolds number motion
is the understanding of the
consequences of hydrodynamic interactions on the collective motion
of an assembly of active components.

The aim of this work is to explore the effect of hydrodynamic
interactions on two rotating dumbbells. Symmetry considerations
forbid a net conformational change over one rotational period for similarly shaped dumbbells with similar couplings to the external field.
One major question explored here is the
influence of differing shapes on the hydrodynamic interactions  between two particles.
Therefore we investigate two interacting
dumbbells of dissimilar shape and coupling to an external rotating field.

We demonstrate by numerical calculations
 that for dissimilar rotating dumbbells the
time reversal symmetry is broken. As a consequence
both dumbbells can attract or
repel each other.
Transitions
between dumbbell attraction and repulsion are induced
by changing the torques or the dumbbell shapes.
Phase diagrams separating regimes of
attraction from regions of repulsion between
the dumbbells are constructed for generic sets of parameters.

We suggest to test our theoretical exploration with
experiments on anisotropic birefringent particles. Such particles
may be rotated
by circularly polarized light \cite{Chaikin:2002.1,Heckenberg:2004.1}.
Another possibility to apply a torque is to place
paramagnetic or ferromagnetic particles in a rotating magnetic field.
Like in nuclear magnetic resonance, magic angle spinning may be used to eliminate time averaged dipole-dipole interactions
that might overshadow the hydrodynamic attraction or repulsion.

\section{Model}\label{sec: model}

\begin{figure}[ht]
\vspace{-3mm}
  \begin{center}
    \includegraphics[width=0.6\columnwidth]{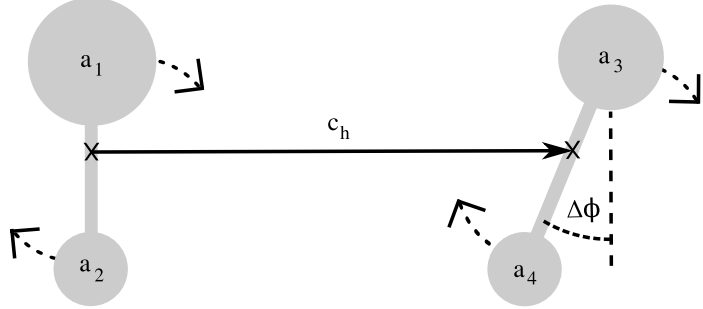}
  \end{center}
\vspace{-5mm}
  \caption{Sketch of two dumbbells with unequal
asymmetries, $a_1 \not =a_3$ and $a_2=a_4$.
The dumbbells are rotated by a driving field  ${\bi f}(t)$, cf.~Eq.~(\ref{fvec}).
$\Delta \phi$ is the angle between the
axes $\hat {\bi r}_{21}$ and $\hat {\bi r}_{43}$ of the dumbbells.
The crosses indicate
the hydrodynamic centers ${\bi c}_{21}$ and ${\bi c}_{43}$ of the
two dumbbells. The distance between them is denoted by ${\bi c}_h$.}
\label{2_dumb_sketch0}
\end{figure}

We investigate the motion of a single dumbbell and of two hydrodynamically
interacting dumbbells in a fluid of viscosity $\eta$ in the low Reynolds number limit.
Due to a torque created by a rotating external field ${\bi f}$
the dummbbells perform a planar rotational motion.
Each dumbbell is composed of two beads which are connected by a bar of constant length as indicated in
figure \ref{2_dumb_sketch0}.

The beads with the effective hydrodynamic radii $a_i$ ($i=1,2,3,4$) at the positions ${\bi r}_i$ ($i=1,2,3,4$)
are described as point particles having drag coefficients $\zeta_i=6\pi \eta a_i$.
The vectors
${\bi r}_{21}={\bi r}_2 - {\bi r}_1$  and ${\bi r}_{43}={\bi r}_4 - {\bi r}_3$
describe the two dumbbell axes, which we assume to be
of constant length $b=|{\bi r}_{21}|=|{\bi r}_{43}|$.
The hydrodynamic centers of the dumbbells,
${\bi c}_{21}$ and ${\bi c}_{43}$,
are given by
\begin{eqnarray}
{\bi c}_{21} ={\bi r}_1 + \frac{a_2}{a_1+a_2} ~ {\bi r}_{21}
\qquad \mbox{and} \qquad {\bi c}_{43} ={\bi r}_3 +
\frac{a_4}{a_3+a_4} ~ {\bi r}_{43}\,.
\end{eqnarray}

The equation of the over-damped motion is
given by
\begin{equation}
 \dot{\vecd{r}}_i = \matr{H}_{ij} \vecd{F}_j\,,
\label{dynamic_equation}
\end{equation}
with  the
  Rotne-Prager mobility matrices for unequal spheres \cite{Prager:1969,Onishi:1984}
\begin{equation}
\matr{H}_{ij} =  \left\lbrace \begin{array}{ll} \frac{1}{6 \pi \eta a_i} \matr{I} & \mathrm{for} \; i=j, \\
\frac{1}{8 \pi \eta r_{ij}} \left[ \left( 1+ \frac{1}{3}\frac{a_i^2+a_j^2}{r_{ij}^2} \right) {\matr{I}} + \left( 1-\frac{a_i^2+a_j^2}{r_{ij}^2} \right) \hat{\vecd{r}}_{ij} \hat{\vecd{r}}_{ij} \right] & \mathrm{for} \; i \neq j, \end{array} \right.\,,
\label{rotdumb:rotne_prager}
\end{equation}
the connection unit vectors
$\hat {\bi r}_{ij} = {\bi r}_{ij}/|{\bi r}_{ij}|={\bi r}_{ij}/ r_{ij}$
and the external forces
$\vecd{F}_j \; (j=1,2,3,4)$ acting on the beads. In our model the rotation of the dumbbells is caused by the potential
\begin{equation}
 V = - V_{21} \hat{\vecd{r}}_{21} \vecd{f} - V_{43} \hat{\vecd{r}}_{43} \vecd{f} \mathrm{,}
\label{potential}
\end{equation}
with the vector
\begin{equation}
\vecd{f} = \left( {\cos (\omega t) \atop \sin (\omega t)} \right)
\label{fvec}
\end{equation}
rotating at a frequency $\omega$ in the $xy$ plane.
The forces ${\bi F}_i= - \nabla_i V$
in Eq.~(\ref{dynamic_equation}) 
can be expressed in terms of the
rotating field ${\bi f}$ as
\begin{equation}
 \vecd{F}_1 = \frac{V_{21}}{r_{21}} \left[ \left( \hat{\vecd{r}}_{21} \times \vecd{f} \right) \times \hat{\vecd{r}}_{21} \right] \qquad \mathrm{and} \qquad \vecd{F}_3 = \frac{V_{43}}{r_{43}} \left[ \left( \hat{\vecd{r}}_{43} \times \vecd{f} \right) \times \hat{\vecd{r}}_{43} \right]\,.
\label{forces}
\end{equation}
The potential in Eq.~(\ref{potential}) yields the following asymmetric relations between the forces
${\bi F}_1 = - \nabla_1 V = \nabla_2 V = -{\bi F}_2$ and ${\bi F}_3 = -\nabla_3 V = \nabla_4 V = - {\bi F}_4$. According to this property together
 with Eq.~(\ref{forces}) the dumbbells are force free. Additionally the forces acting on the individual beads are perpendicular to the dumbbell axes,
which results in a rotation
of the dumbbells.

The form of the potential $V$ for the driving
field is similar to that of
magnetic dipoles in an external magnetic field where the interaction between the
dipoles is neglected. But other means to propel the dumbbells are possible.
For example, non-spherical birefringent
particles can be rotated by circularly polarized light.
The magnitude of the
torques imposed on the dumbbells can easily be tuned by the applied
power of the light \cite{Chaikin:2002.1,Heckenberg:2004.1}.

\section{Dumbbell dynamics}\label{sec: results}

The balance between the driving and the viscous torques leads to
phase angles
$\phi_{21}$ and $\phi_{43}$
between the dumbbell orientations ${\bi r}_{ij}$
 and the driving field  ${\bi f}(t)$
with $\cos(\phi_{ij})= {\bi r}_{ij}\cdot
{\bi f}/|{\bi r}_{ij}\cdot {\bi f}|$. The dynamics of the dumbbells is analyzed in this section.
The conditions under which
both dumbbells attract
or repel are discussed.

\subsection{Single dumbbell}

The two beads of a single dumbbell ($V_{43} = 0$ and $V_{21} = V_0$) move on circular trajectories
around  their hydrodynamic center
$\vecd{c}_{21}$. The radii $R_{1,2}$ of those trajectories depend on
the effective hydrodynamic radii $a_{1,2}$ of the two beads and the length $b$
of the dumbbell axis. In the Oseen approximation they are given by
\begin{eqnarray}
R_1  =  \frac{a_2~b}{a_1+a_2}  \qquad \mathrm{and} \qquad R_2  =  \frac{a_1~b}{a_1+a_2} ~~ \mathrm{.}
\label{radii_of_bead_motion}
\end{eqnarray}

For a sufficiently strong external field the dumbbell axis ${\bi r}_{21}$ rotates
synchronously with the driving field ${\bi f}(t)$ at the same frequency $\Omega=\omega$ after a transient period.
The stationary phase angle
$\phi_{21}$ is given by the expression
\begin{equation}
 \sin \phi_{21} = - \frac{\pi \eta \omega b^2}{V_0 \left[ \displaystyle{\frac{1}{6 a_1} + \frac{1}{6 a_2} - \frac{1}{4 b} \left( 1+ \frac{a_1^2+a_2^2}{3b^2} \right)} \right]} ~\mathrm{.}
\label{phase_shift}
\end{equation}
$\phi_{21}$ tends to zero for decreasing values of
the ratio $\omega/V_0$, i. e. either for a decreasing frequency or
for increasing values of the coupling  $V_0$.
If the expression
on the right hand side in Eq.~(\ref{phase_shift})
takes values outside of the
interval $[-1,1]$, the driving torque is too small to enforce a synchronous rotation.
These analytical results are in agreement with
numerical integrations of Eq.~(\ref{dynamic_equation}).

\subsection{Two dumbbells}

\begin{figure}[ht]
  \begin{center}
   \includegraphics[width=0.3\columnwidth, angle=-90]{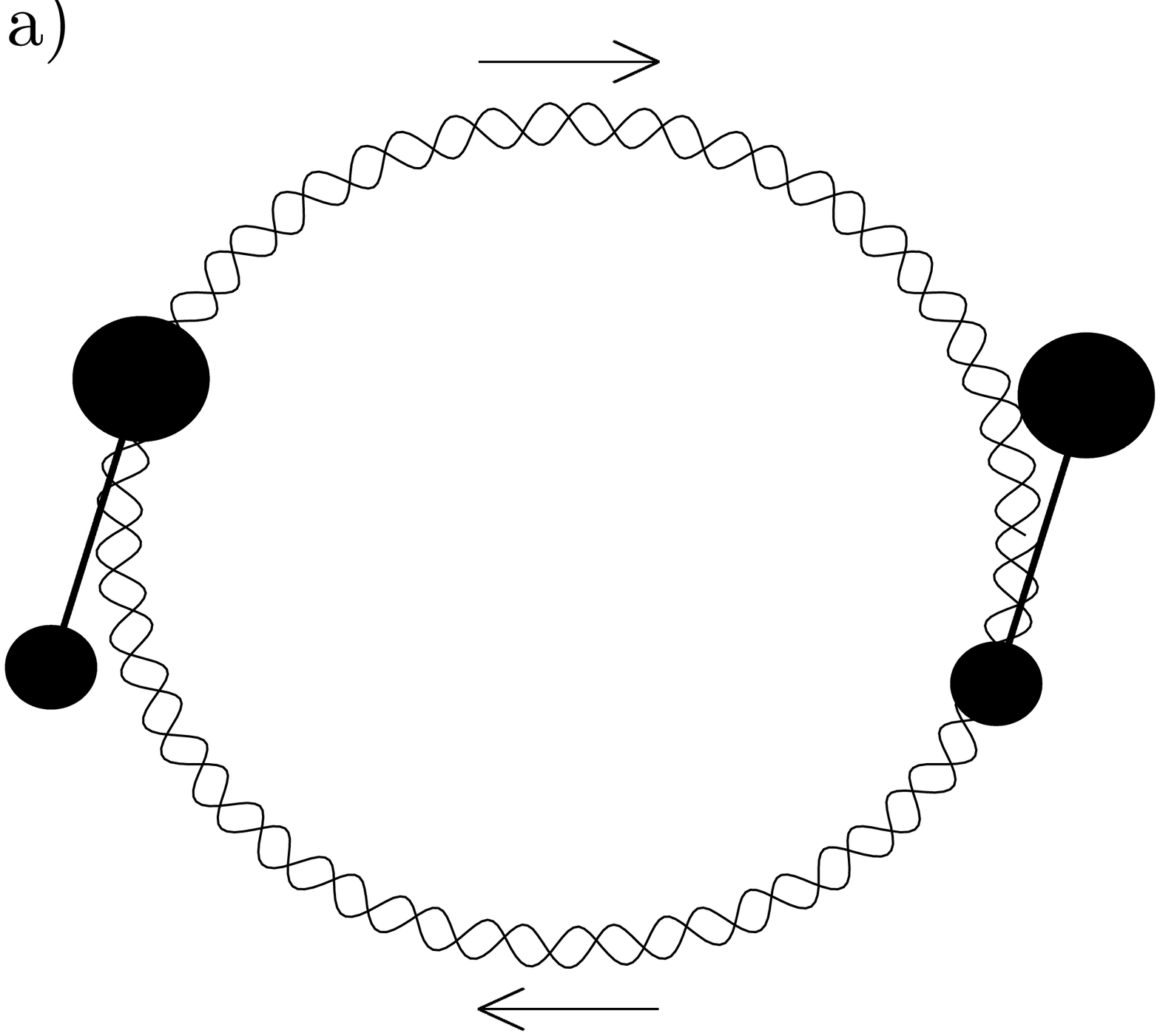}
   \hspace{1cm}
   \includegraphics[width=0.3\columnwidth, angle=-90]{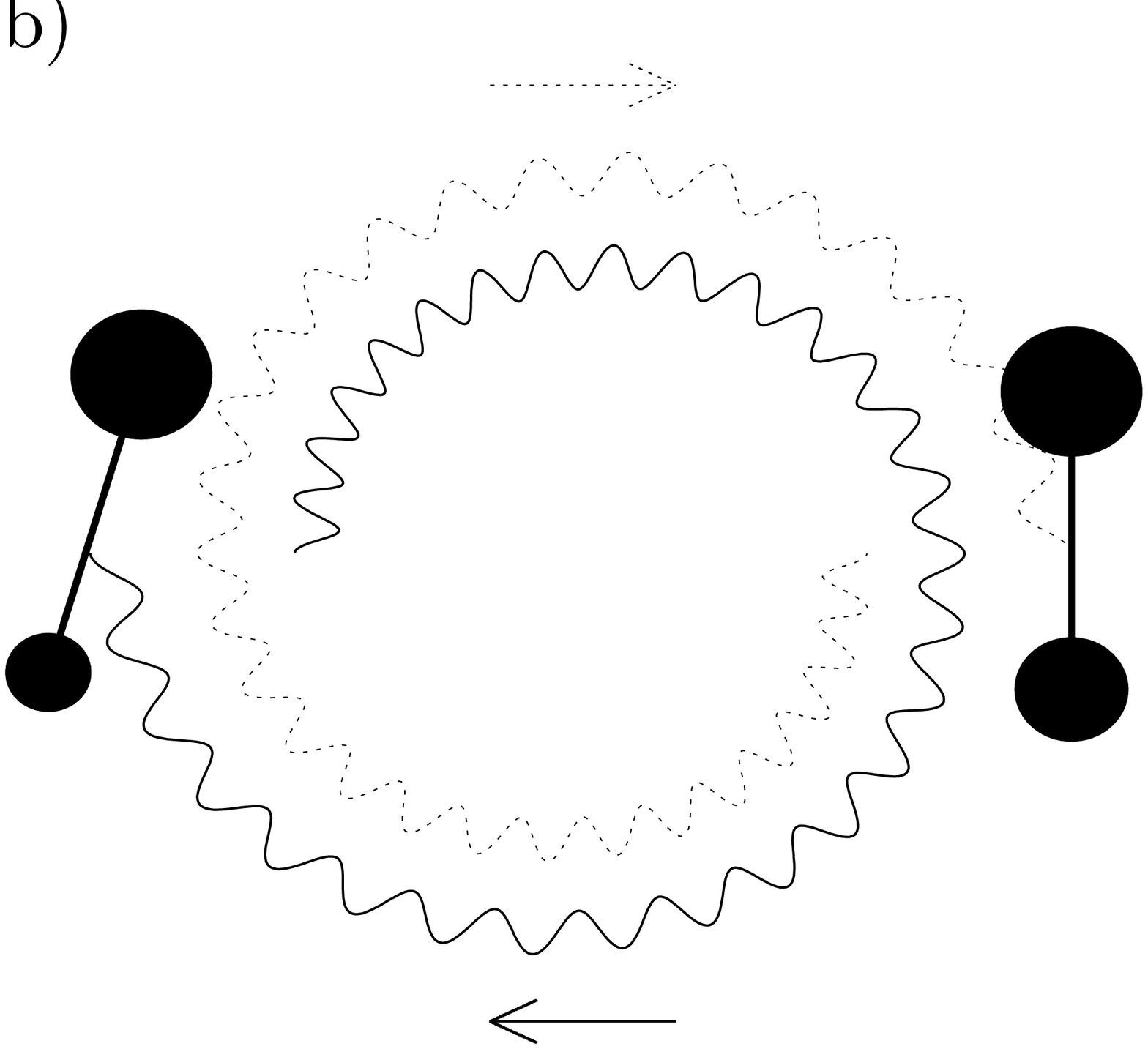}
  \end{center}
  \caption{In part (a) the trajectories of the centers of two rotating,
equally oriented and equally asymmetrically shaped dumbbells are sketched.
In addition to that the coupling parameters, $V_{21}=V_{43}$, are identical.
The trajectories of both hydrodynamic
centers ${\bi c}_{21}$ and ${\bi c}_{43}$ have
the same mean radius, which corresponds to the
mean length of the vector $\bi{c}_h$.
Part (b) corresponds to the case of two dumbbells with differently asymmetric shapes
and different coupling strengths ($V_{21} \neq V_{43}$). The resulting
trajectories are open and the mean distance $D(t) =|\bi{c}_h |$ increases during the motion.
}
\label{2_dumb_sketch1}
\end{figure}

In the case of two rotated dumbbells each one experiences
 the perturbed liquid flow created by the respective other one. Due to this
interaction the hydrodynamic centers  ${\bi c}_{ij}$
are set in motion.
In the mean this results in a circular or a spiral like motion
as indicated in figure \ref{2_dumb_sketch1}.
In the following considerations we fix the effective radii of one bead of each dumbbell at the
value $a_2=a_4=1$ and the length of the axes of both dumbbells at $b=3$ for simplicity.
The (asymmetric) shapes of the dumbbells and the 
couplings to the external field can be adjusted independently by
varying the bead radii $a_1$ and $a_3$ as well as the coupling parameters
$V_{21}$ and $V_{43}$.
The viscosity $\eta$, the angular frequency $\omega$,
the coupling $V_{ij}$ and the length $b$ form a dimensionless group $\eta\omega b^3/V_{ij}$. It is therefore sufficient to vary the coupling and keep the viscosity $\eta=10$ and the angular frequency $\omega=5\times 10^{-4}$ fixed.

The complete conformation of the system (up to a rotation of the whole system)
can be described by the distance $D(t)$
between the hydrodynamic centers of the dumbbells,
\begin{equation}
 D(t) = |{\bi c}_h| = | {\bi c}_{21}-{\bi c}_{43} | = \left| {\bi r_1} +\frac{a_2}{a_1+a_2}~{\bi r}_{21} -{\bi r}_3 -\frac{a_4}{a_3+a_4}~{\bi r}_{43} \right|\,,
\end{equation}
as well as the conformation angles $\psi_{21}$ and $\psi_{43}$, which are given by
\begin{equation}
  \cos(\psi_{ij}) = \frac{\bi{c}_h \cdot {\bi r}_{ij}}{|\bi{c}_h \cdot {\bi r}_{ij}|} \,.
\end{equation}
The hydrodynamic interactions between the dumbbells
induce a slow rotation of ${\bi c}_h$ and
oscillations of D(t) around some mean distance
as indicated by figure \ref{2_dumb_sketch1}.
The frequency $\Omega'$ of these oscillations is
the rotation frequency of the dumbbell axes ${\bi r}_{21}$ and ${\bi r}_{43}$
with respect to $\bi{c}_h$. Therefore the frequency of the slow rotation of $\bi{c}_h$ is
equal to the difference $\Omega-\Omega'$.

\subsubsection{Two dumbbells having identical shapes and couplings.}

The dumbbell dynamics is considerably simplified for
equally asymmetric shapes of the two dumbbells, $a_1=a_3$, and equal
couplings, $V_{21}=V_{43}$.
If the two dumbbells are also
equally oriented, as sketched in figure \ref{2_dumb_sketch1}(a), the phase angles are equal,
$\phi_{21} = \phi_{43} = \phi$,
and the conformation angles are identical for all times, $\psi_{21}=\psi_{43}=\psi$.
For two oppositely oriented
dumbbells the phase difference has the constant value $\Delta \phi = \pi$ whereas the conformation angles add up to $\psi_{21}+\psi_{43}=\pi$.

\begin{figure}[ht]
\vspace{-3mm}
  \begin{center}
    \includegraphics[width=0.45\columnwidth,angle=-90]{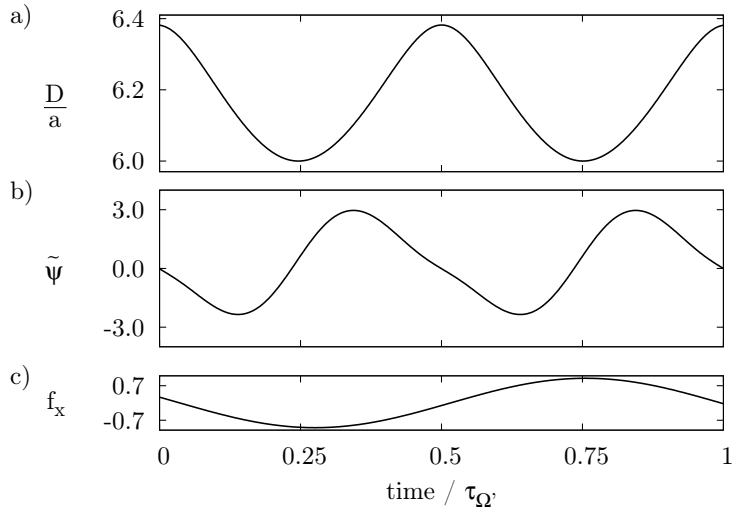}
  \end{center}
\vspace{-5mm}
  \caption{Part (a) shows the dimensionless distance $D(t)/a$
 for two equally asymmetrically shaped and equally oriented dumbbells
with  $a_1=a_3=1.2$ and $V_{21} = V_{43}=0.7$.
In part (b) we present the periodic contribution to
the phase angle $\tilde \psi =\psi - \Omega' t$. For comparison
the
periodic variation of the $x$ component of the driving field ${\bi f}$ is shown in part (c).
}
\label{conformation_versus_time}
\end{figure}

According to the symmetry of the system,
the distance $D(t)$ between the hydrodynamic centers
of two equally oriented dumbbells
oscillates around a
mean distance with the frequency $2\Omega'$ as indicated by the sketch in figure \ref{2_dumb_sketch1}(a).
The corresponding
oscillation period is $\pi/\Omega'$,
which is shown for a specific parameter set in
figure \ref{conformation_versus_time}(a). In contrast to this,
the distance $D(t)$ for
 two dumbbells
with opposite shape orientation oscillates with a
period of $\tau_{\Omega'}=2\pi/\Omega'$ being twice as long as in the former case.
For comparison we show
the $x$ component of the driving field $f_x(t)$ in figure \ref{conformation_versus_time}(c),
which illustrates the inequality of the frequencies,
 $\Omega' < \Omega$, and thus the inequality of the oscillation periods $\tau_{\Omega'}>\tau_{\Omega}=2\pi/\Omega$.

In figure \ref{conformation_versus_time}(b) the
oscillatory contribution
$\tilde \psi$  to the phase
$\psi=\Omega't + \tilde \psi$ is shown. Since the coupling strengths of
the dumbbells are constant, the
oscillation of $\psi$ about a linearly growing
part confirms that the rotational
hydrodynamic drag changes as a function
of the relative orientation of the dumbbell
axes with respect to ${\bi c}_h$.
In the case of equally oriented dumbbells
the viscous drag is maximal in the ranges about the conformation angles
$\psi=(2n-1)\pi/4$ ($n=1,2,3,4$)
where fluid between the dumbbells is squeezed out or
sucked into the region between the dumbbells.
These are also the phases of the rotation during which the repulsion (resp. attraction)
between the dumbbells is strongest.

According to the symmetry of the system the attractive forces acting
between the two dumbbells in one quarter of a period ($0 < t < \tau_{\Omega'}/4$)
are compensated during the consecutive quarter ($\tau_{\Omega'}/4 < t < \tau_{\Omega'}/2$)
by repulsive forces of the same magnitude.
This causes an oscillation
of $D(t)$ with the frequency $2 \Omega'$ around
some mean value, but without a net attraction or repulsion.
In the case of oppositely oriented dumbbells two consecutive
maxima of the viscous drag have different amplitudes and therefore
the attractive and repulsive hydrodynamic forces do not cancel each other completely
during one half of an oscillation period but during
a whole rotation with the period $\tau_{\Omega'}$.

During the rotation the viscous drag acting on the
inner beads, which are closer to the respective other dumbbell,
is higher than the drag on the outer beads of the dumbbells.
These differing drag forces induce an oscillation of
the phase angles $\phi_{ij}$ during the rotation.

In the ranges of increasing (decreasing) viscous drag the rotation frequency $\Omega'$ of
the dumbbell axes decreases (increases). When $\Omega'$ is larger (smaller) than $\omega$
the phase delays $\phi_{ij}$ of the dumbbells decrease (increase).

\begin{figure}[ht]
\vspace{-3mm}
  \begin{center}
    \includegraphics[width=0.32\columnwidth,angle=-90]{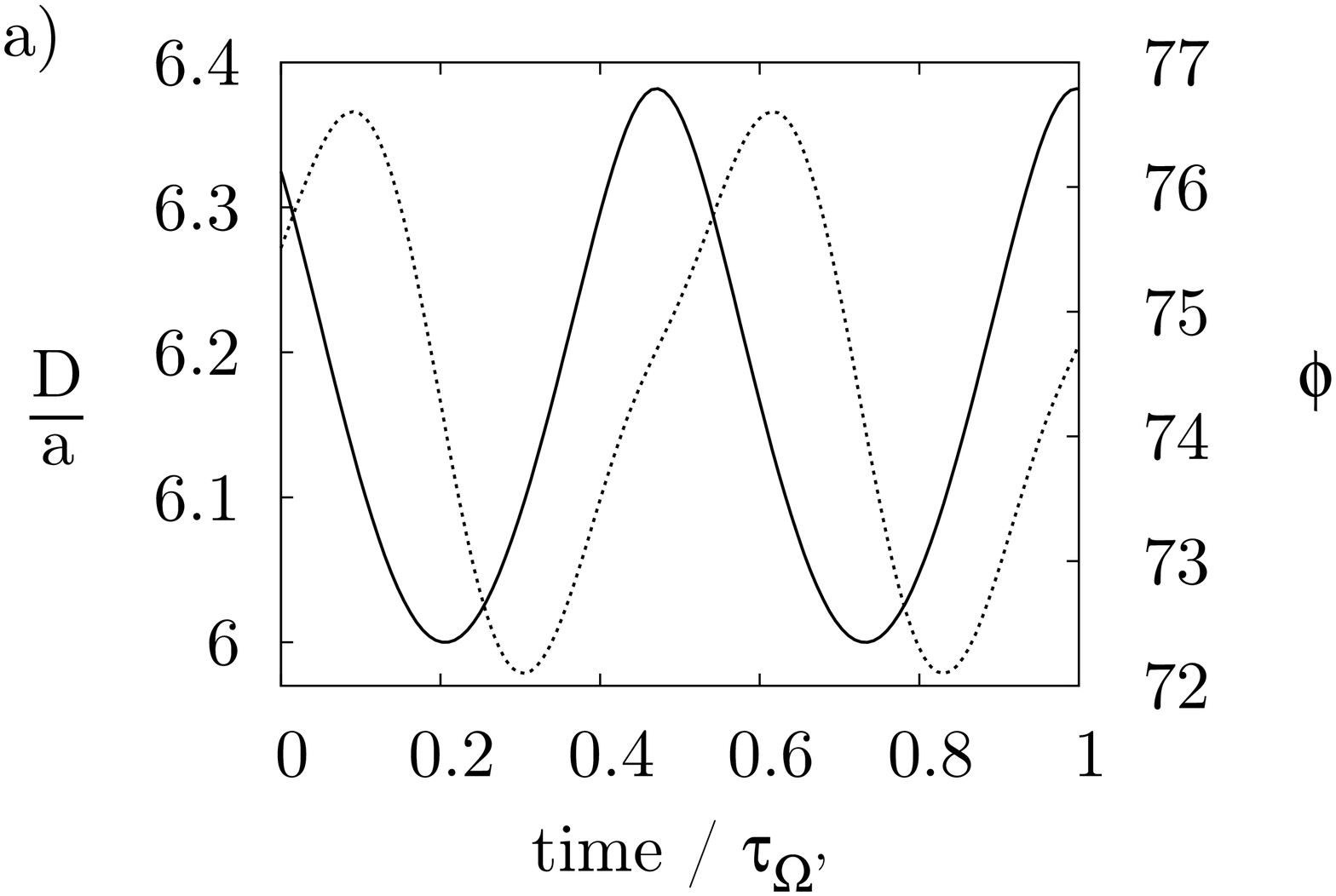}
     \hspace{1mm}
    \includegraphics[width=0.32\columnwidth,angle=-90]{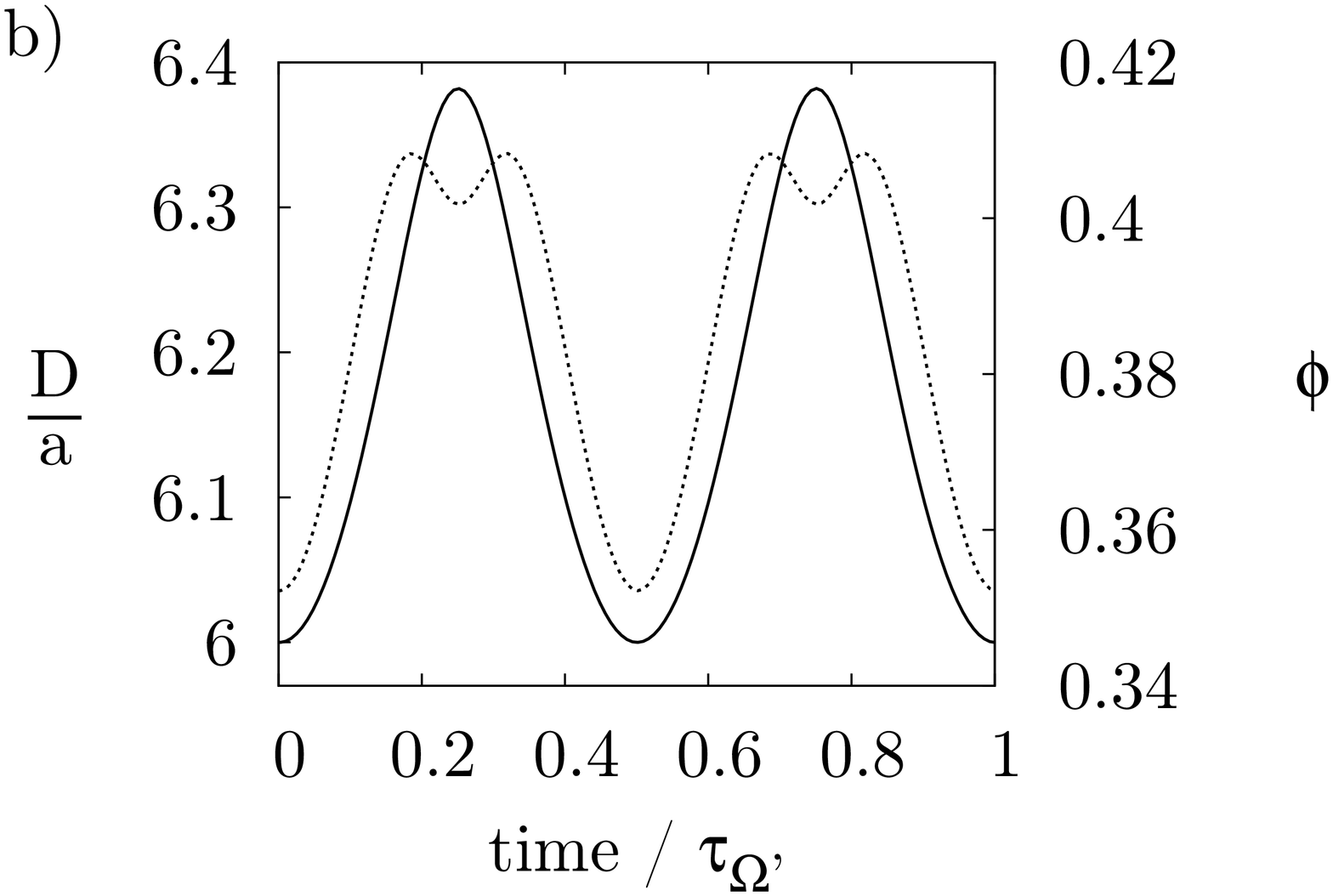}
  \end{center}
\vspace{-5mm}
  \caption{The distance $D(t)$  (solid line)
and the phase lag $\phi=\phi_{21}=\phi_{43}$ (dashed line)
 are shown as a function of time
for $a_1=a_3=1.2$. In part (a)
the coupling is weak, $V_{21}=V_{43}=0.7$, and in part (b)
the coupling constants are large, $V_{21}=V_{43}=100$.
}
\label{phase_lag_weak_and_strong}
\end{figure}

The oscillations of $D(t)$ and $\phi(t)$ are shown as a function of time
for two different values of the coupling constants in figure \ref{phase_lag_weak_and_strong}.
In Part (a) the coupling
to the driving field is weak,
$V_{21}=V_{43}=0.7$, and in part (b) the
coupling is strong, $V_{21}=V_{43}=100$.
The amplitudes of the oscillations of the distance are the same in both cases.
Is contrast to that the phase lag $\phi$
of the dumbbell axes with respect to
the orientation of the external field is smaller
in the case of large values of the coupling.
Moreover the diagrams in figure \ref{phase_lag_weak_and_strong}
show that the oscillations of $\phi$
are in phase with the oscillations of the distance
in the case of strong couplings to the external field.
Contrary to that $\phi(t)$ is delayed with respect to $D(t)$ for weak couplings.

The reason for this delay is the following:
While the distance between the dumbbells increases
the drag on the dumbbells also grows. So the rotation frequency $\Omega'$
of the dumbbells must decrease. As soon as it is below $\omega$ the phase angle
$\phi$ grows and so does the torque on the dumbbells ($\sim \sin\phi$). But for weak couplings the
torque builds up much slower than for strong couplings. So for small coupling constants even in the beginning of
the domain where the distance (and thus the drag) decreases $\phi$ is still
growing because the torque is too weak.
At some point the torque is large enough to overcome the drag so that $\phi$
decreases again afterwards.
An analogous argument holds in the ranges where $D(t)$ is increasing.
The shift between the curves of $D(t)$ and $\phi(t)$ has to be smaller than
a quarter of an oscillation period in our considerations
because otherwise the field ${\bi f}(t)$
would be too weak to enforce a synchronous rotation of the dumbbells at all.

For strong couplings there are small dips at the maxima of $\phi(t)$. These correspond to local minima of the drag, which
occur when the dumbbells are aligned with each other so that the four beads lie
on a single line. In this conformation the distance between the dumbbells is maximal.
An equivalent phenomenon can be seen in the plot of $\phi(t)$ for weak coupling.
Of course there is also a local minimum of the drag when the dumbbells are aligned with each other and the distance is maximal.
But due to the weak torque acting at that stage there is no dip but only a slight flattening of the curve in this case.

\begin{figure}[ht]
\vspace{-3mm}
  \begin{center}
    \includegraphics[width=0.355\columnwidth,angle=-90]{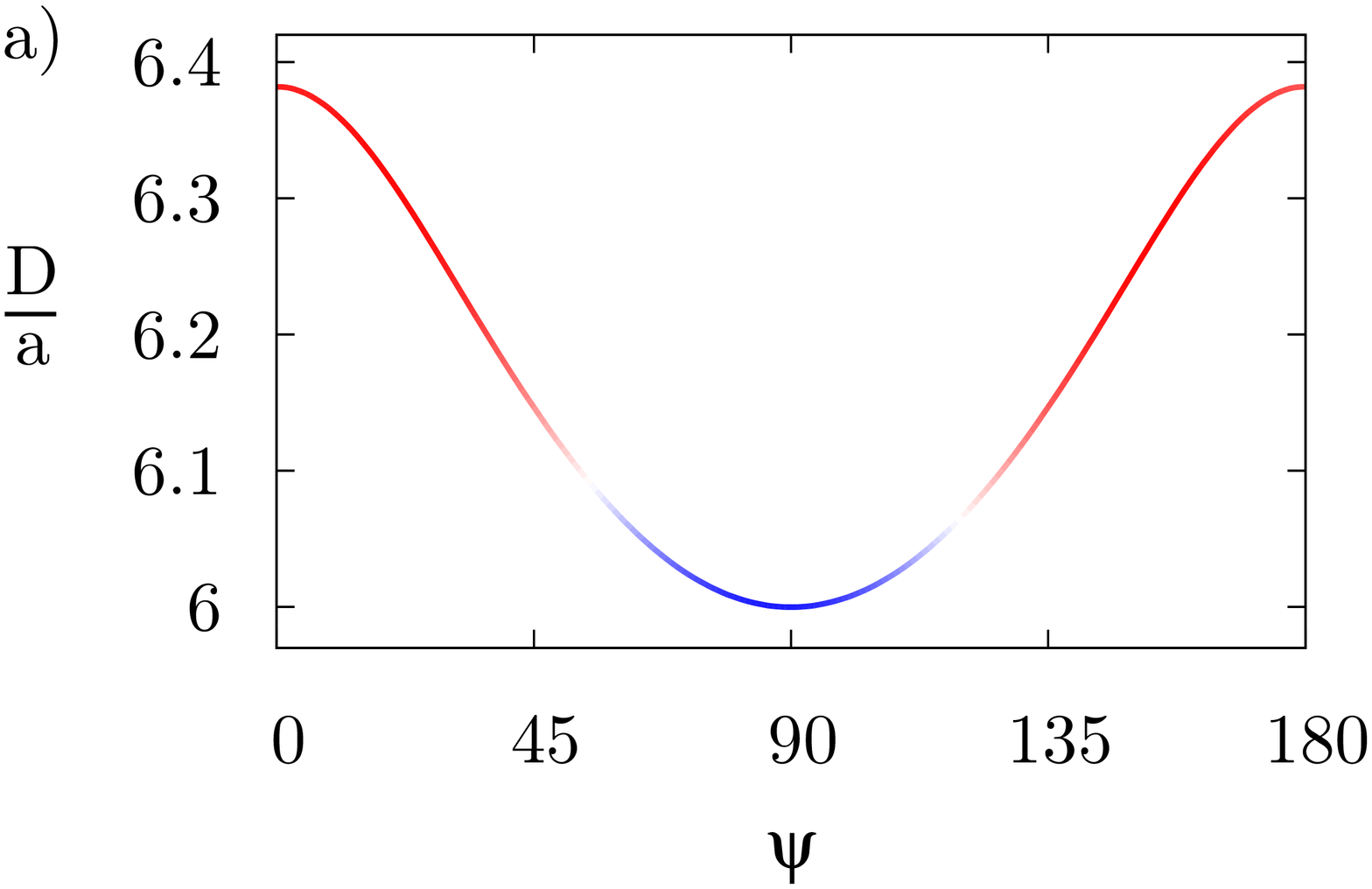}
     \hspace{-5mm}
    \includegraphics[width=0.355\columnwidth,angle=-90]{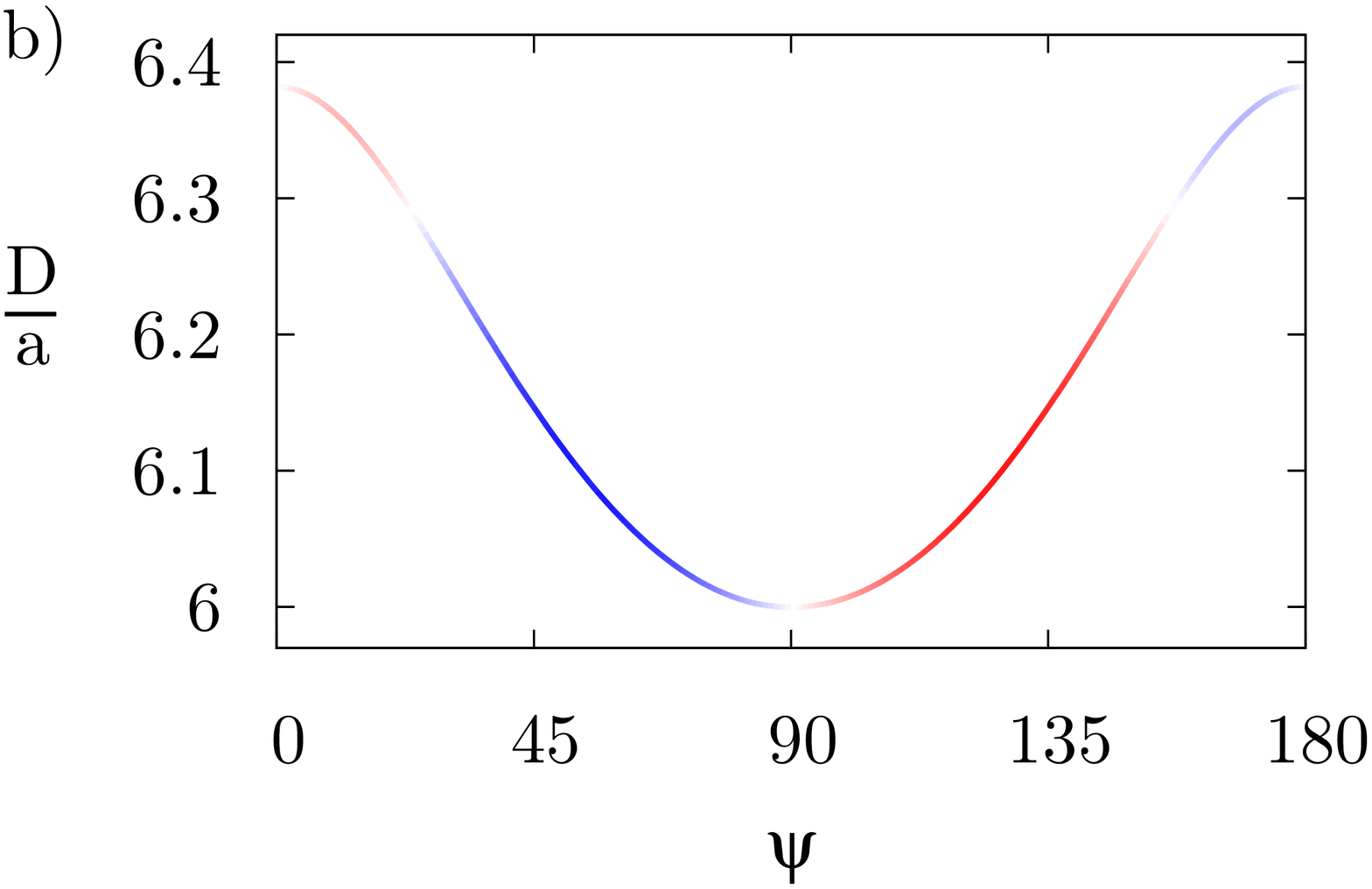}
  \end{center}
\vspace{-5mm}
  \caption{
Trajectories in the $D-\psi$ plane for a system of two dumbbells with $a_1=a_3=1.2$
and equal coupling constants. In part (a) the couplings are $V_{12}=V_{34}=0.7$ and
in part (b) $V_{12}=V_{34}=100$.
Both trajectories are invariant under the transformation $\psi \rightarrow -\psi$
showing the reciprocity of the motion. What is more the trajectories fall on top of each other,
which illustrates the universality of the trajectories. However, according
to the two different couplings the
rates $\dot\psi$ along the trajectories differ. The rates $\dot \psi$
 are color coded where red denotes a rate below the average one and blue a rate above.}
\label{D_versus_psi}
\end{figure}

In the coordinate system rotating with the connection vector ${\bi c}_h$
the conformation after one period $\tau_{\Omega'}$ maps on top of the original conformation.
The reciprocity of the motion is illustrated in
figure \ref{D_versus_psi}, where
the distance $D$ is plotted as a function
of the conformation angle
$\psi$ for weak couplings, $V_{12}=V_{34}=0.7$, in part a)
and strong couplings, $V_{12}=V_{34}=100$, in part b). Both curves are invariant
under the transformation $\psi \rightarrow -\psi$. This underlines the reciprocity of the motion
as it can already be seen in the plot of $D(t)$ which is invariant under time reversal $t \rightarrow -t$.
In addition to that the curves for weak and strong couplings are identical in the conformation space $D-\psi$
as shown in figure \ref{D_versus_psi}. In fact the trajectories shown are universal for all values of the coupling constants
as long as they are large enough so that the dumbbells can follow the field ${\bi f}(t)$ synchronously.

However,  the rate of change of the conformation angle $\dot\psi$ along the curves
depends very much on the coupling parameters (cf. figure \ref{phase_lag_weak_and_strong}). This is indicated by the
color code along the two
curves, where red and blue mark the rates $\dot\psi$
below and above the angular frequency $\Omega'$.
This shows that irrespective of the strength of the
driving field the dumbbells pass through the
same set of conformations during one cycle, although each conformation
is passed at a different rate, as indicated by the distribution of
the different colors along both lines.
The reciprocity of the motion ensures that for a given
rotational frequency the distance of the dumbbells described by
the Stokesian dynamics is locked to the individual orientation of the dumbbells with respect to ${\bi c}_h$
 and returns to the same value after one period.

\subsubsection{\label{sec222} Two dumbbells of dissimilar shapes or couplings.}

\begin{figure}[ht]
\vspace{3mm}
  \begin{center}
   \includegraphics[width=0.33\columnwidth, angle=-90]{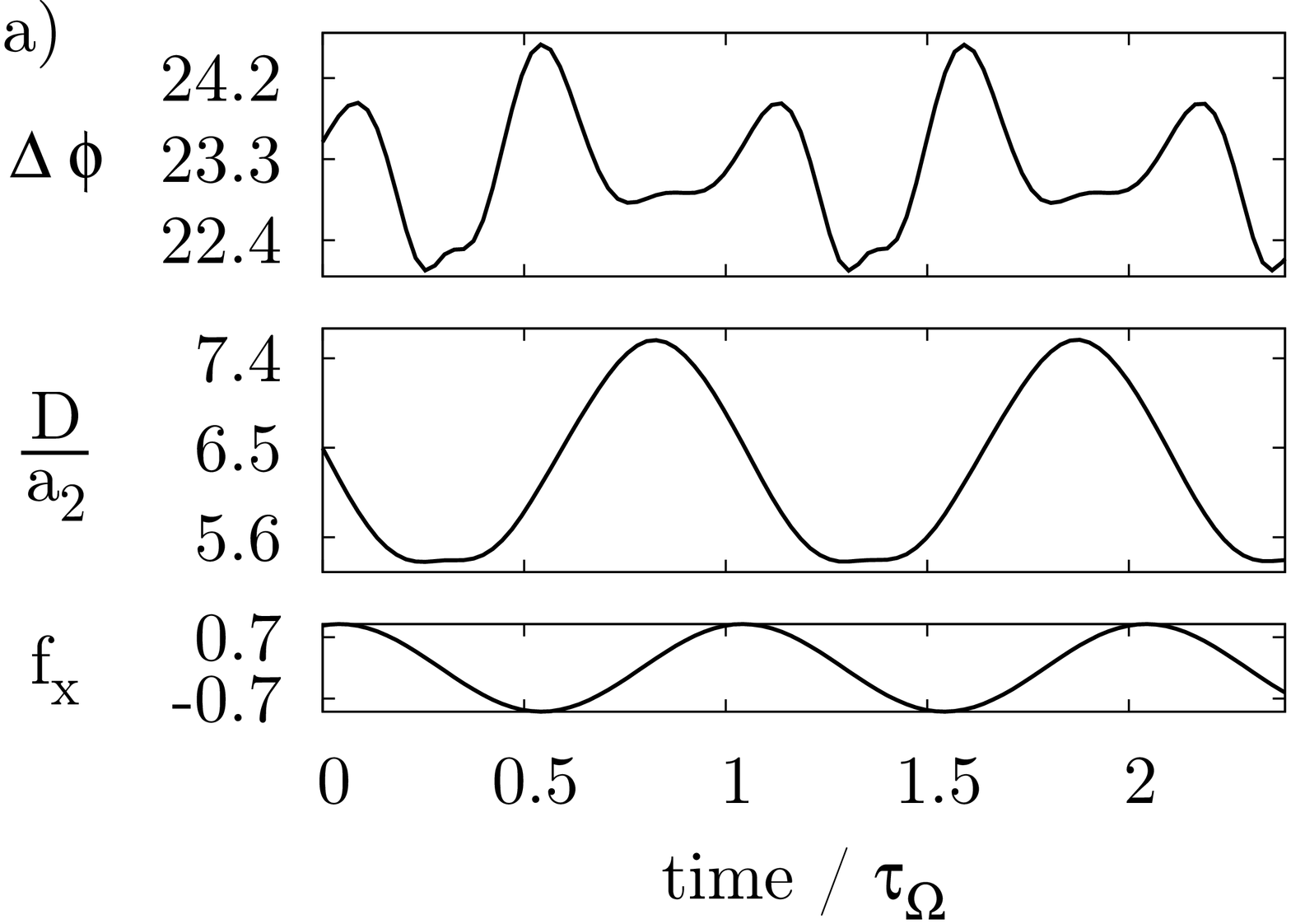}
   \hspace{2mm}
   \includegraphics[width=0.33\columnwidth, angle=-90]{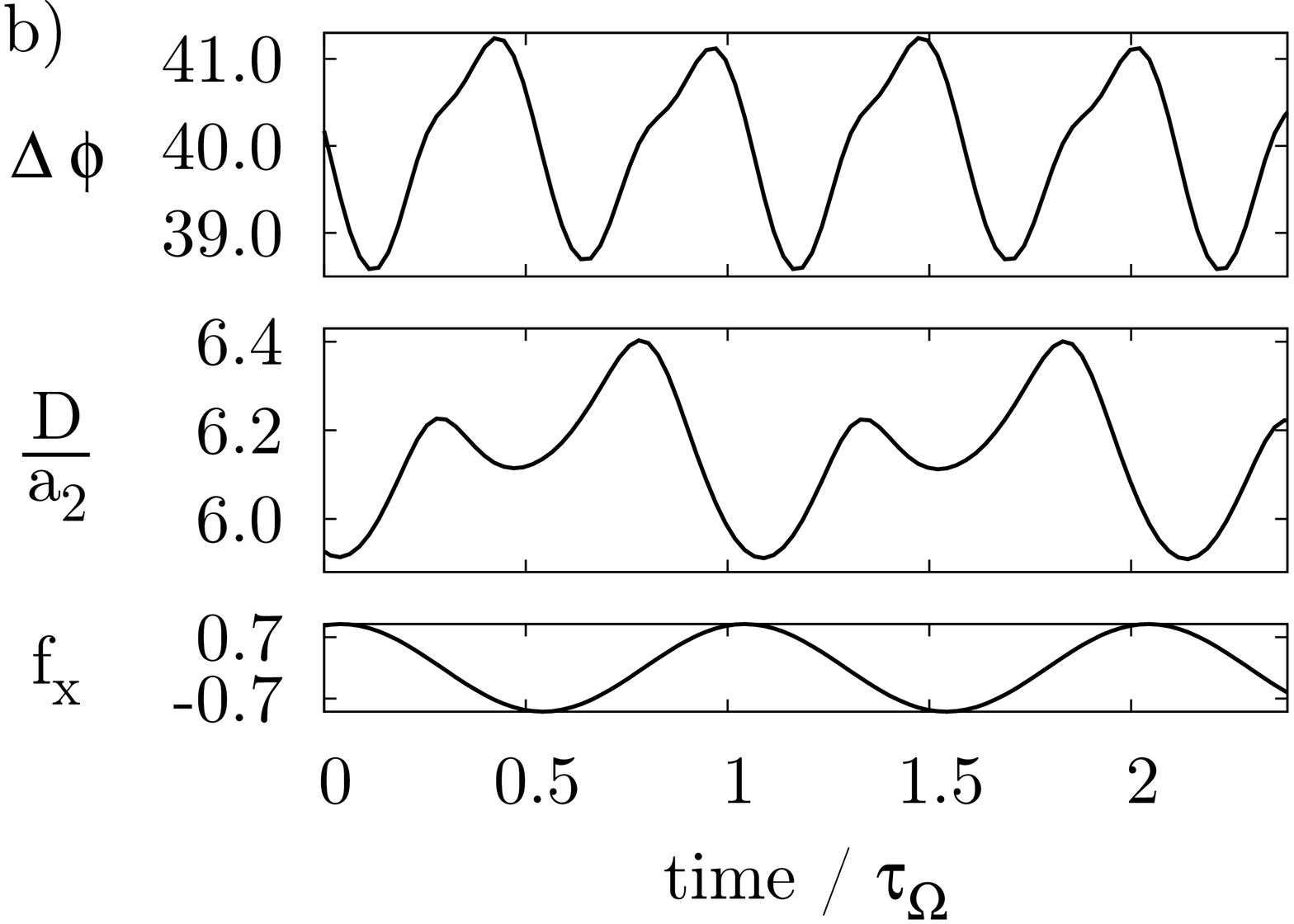}
  \end{center}
\vspace{-5mm}
  \caption{The upper parts show the phase difference $\Delta \phi$ between the two axes of the rotating
dumbbells as a function of time.
The distance $D(t)$ is plotted in the middle parts
and the lower parts show the $x$ component of the driving field $\bi{f}$.
In part (a) the two dumbbells repel each other. The corresponding parameters
are $a_1 = 1.8$, $a_3 = 0.6$, $V_{21} = 1.2$ and $V_{43} = 1.0$.  In part (b) the
dumbbells attract each other. There the parameters are
$a_1 = 1.2$, $a_3 = 1.1$, $V_{21} = 3.2$ and $V_{43} = 0.8$. }
\label{ps_2_dumb}
\end{figure}

For two differently shaped dumbbells or for different couplings to the external field  the phase difference $\Delta \phi$ and
the distance  $D(t)$
are plotted as functions of the time in figure \ref{ps_2_dumb} for two different sets of
parameters. For comparison also
the $x$-component $f_x$ of the driving field is
shown. In contrast to the previous section
the phase difference $\Delta \phi(t)$ oscillates
in time. The time dependence is different for different parameters
as indicated in  figure \ref{ps_2_dumb}(a) for the
parameters $a_1 = 1.8$, $a_3 = 1.0$, $V_{21} = 1.2$, $V_{43} = 1.0$
and in figure \ref{ps_2_dumb}(b) for the parameters $a_1 = 1.2$, $a_3 = 1.1$, $V_{21} = 1.0$, $V_{43} = 2.0$.

In the previous section
$D(t)$ was either oscillating with the frequency $2\Omega'$ or
$\Omega'$, but in both cases $D(t)$ was symmetric with respect to a
reflection of time, cf.  figure \ref{phase_lag_weak_and_strong}.
Here both, $D(t)$ and $\Delta \phi(t)$,
are not symmetric anymore
with respect to time reflection.

\begin{figure}[htb]
\vspace{-3mm}
  \begin{center}
    \includegraphics[width=0.33\columnwidth,angle=-90]{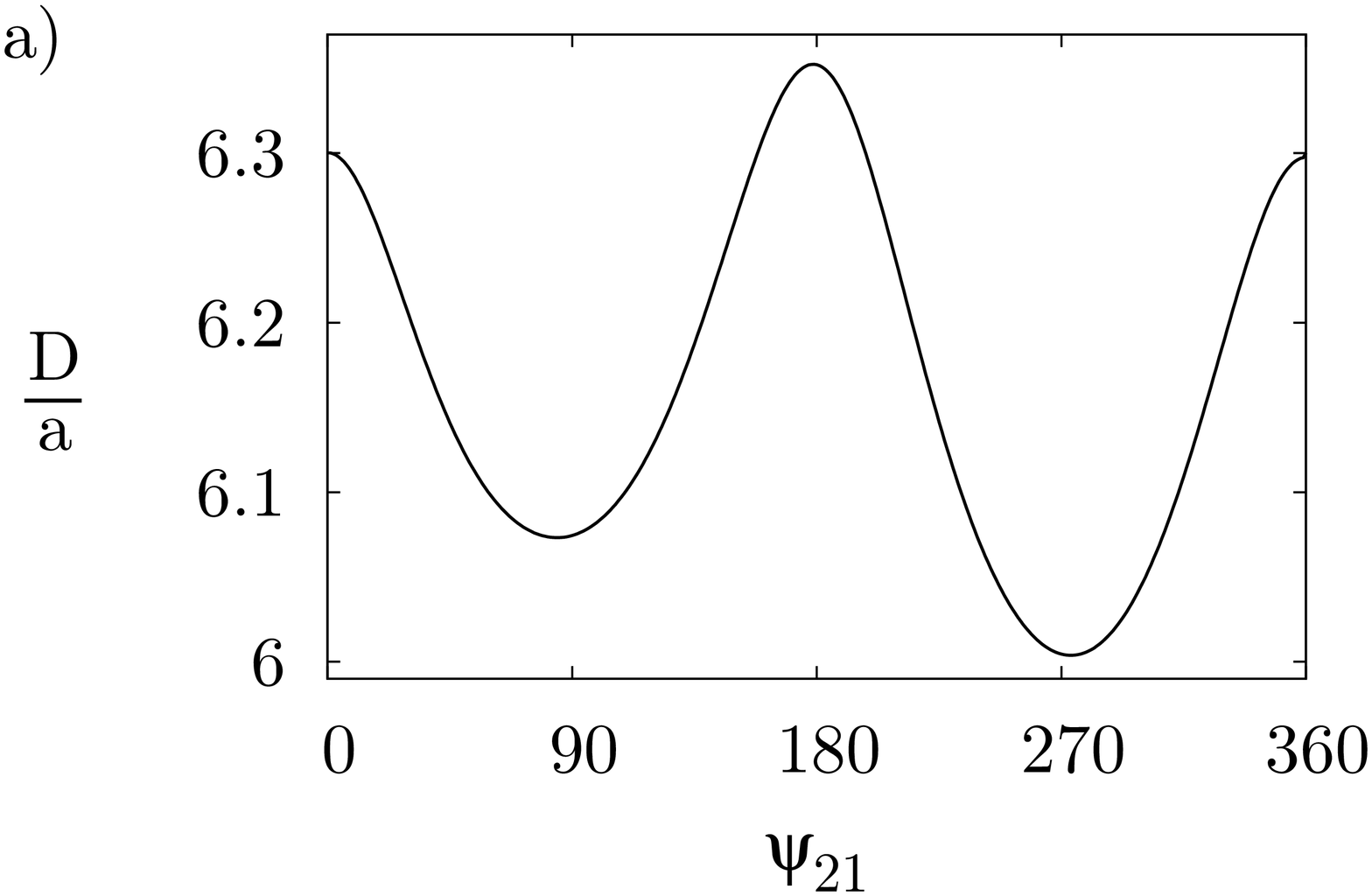}
     \hspace{2mm}
    \includegraphics[width=0.33\columnwidth,angle=-90]{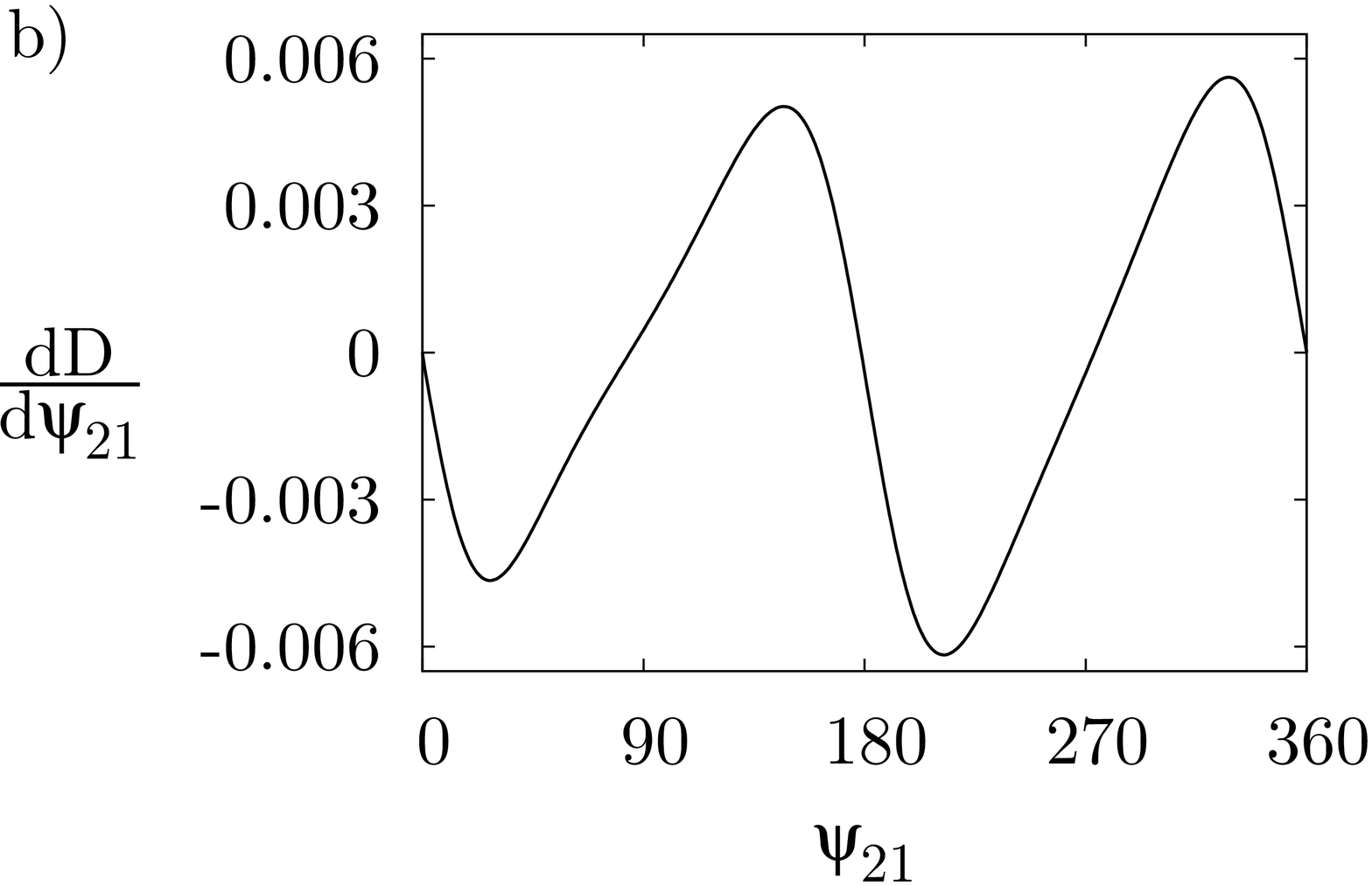}
  \end{center}
\vspace{-5mm}
  \caption{In part (a) the distance $D(t)$ between two dumbbells is plotted as a function of the conformation angle $\psi_{21}$ between the axis of the first dumbbell and the vector $\bi{c}_h$.
Part (b) shows the derivative of $D(t)$ as a function of $\psi_{21}$. The parameters are $a_1=1.2$, $a_3=1.1$, $V_{21}=3.2$, and $V_{43}=0.8$.
The dynamics of this system is much more complex than for equally asymmetric dumbbells with equal couplings (cf. figure \ref{D_versus_psi}).
 }
\label{distdumbbell}
\end{figure}

\begin{figure}[htb]
\vspace{-3mm}
  \begin{center}
    \includegraphics[width=0.33\columnwidth,angle=-90]{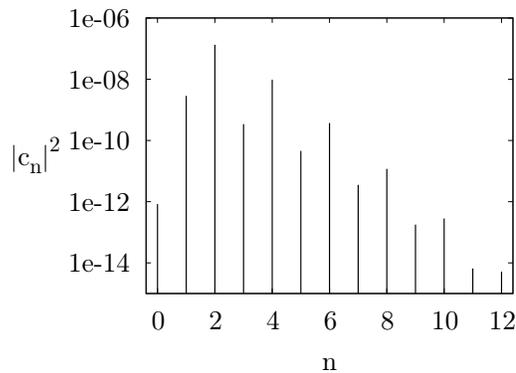}
  \end{center}
\vspace{-5mm}
  \caption{The Fourier modes of the function $\frac{\partial D}{\partial \psi_{21}}$ are shown. There is a non-vanishing zeroth order coefficient that leads to an attraction or a repulsion between the dumbbells. However, there are eight higher order coefficients that have larger values which indicate a quite complex motion.}
\label{fourierpsi}
\end{figure}

As in the case of equally asymmetric dumbbells the
viscous drag is a function of the angles which the
dumbbell axes enclose with the vector
${\bi c}_h$.
There are also ranges of the phase angles,
in which liquid is squeezed out between
the dumbbells and others where fluid is sucked in. As a consequence
the length of the vector ${\bi c}_h$ is also
oscillating here, i. e. there are phase ranges of
dumbbell attraction, which are followed by phases of dumbbell repulsion and so on.

For finite values of
$\Delta \phi  \not =0$, which are a consequence of the
broken symmetry, the conformation angles $\psi_{21}$ and $\psi_{43}$ differ from each other. Combined with
the hydrodynamic interactions, which depend nonlinearly on the
distances between the beads, the magnitude and the directions of
the forces between the beads during one
rotation cycle are quite complex. Therefore the
trajectories in the configurational space
are not universal anymore.

For one parameter set the distance D(t) is plotted as a function of the
conformation angle $\psi_{21}$ in figure \ref{distdumbbell}(a).
It is clear from first sight that the behavior is much more complex here than
in the case of equally asymmetric dumbbells with equal couplings, which was shown in figure \ref{D_versus_psi}.
Especially the magnitudes of the attraction and repulsion between the dumbbells during the different
stages of the motion are different here.
So a priori it is not clear anymore
whether the dumbbells can
attract or repel each other during a whole rotational period. In order to illustrate
 the complex behavior of $D$ as a function of $\psi_{21}$ in more detail, figure \ref{distdumbbell}(b)
shows the derivative $\frac{\partial D}{\partial \psi_{21}}(\psi_{21})$. In the ranges of the conformation angle $\psi_{21}$
where this function is positive the dumbbells repel each other and when the function is negative they attract each other.
The Fourier modes of this function are given in figure \ref{fourierpsi}. In this spectrum one can clearly see that there
is a non-vanishing coefficient of zeroth order which means that the distance between the dumbbells changes
during a complete dumbbell rotation. But it is also obvious that this net repulsive or attractive effect
is superimposed by much stronger oscillations. So it is not possible to determine in which phase of the motion
the decisive effect takes place that causes the overall repulsion or attraction. In fact all coefficients from $n=1$ $(\sim \e^{i\psi_{21}})$ up to $n=8$ $(\sim \e^{8i\psi_{21}})$
are larger than the zero order coefficient.

\begin{figure}[ht]
\vspace{-3mm}
  \begin{center}
    \includegraphics[width=0.34\columnwidth, angle=-90]{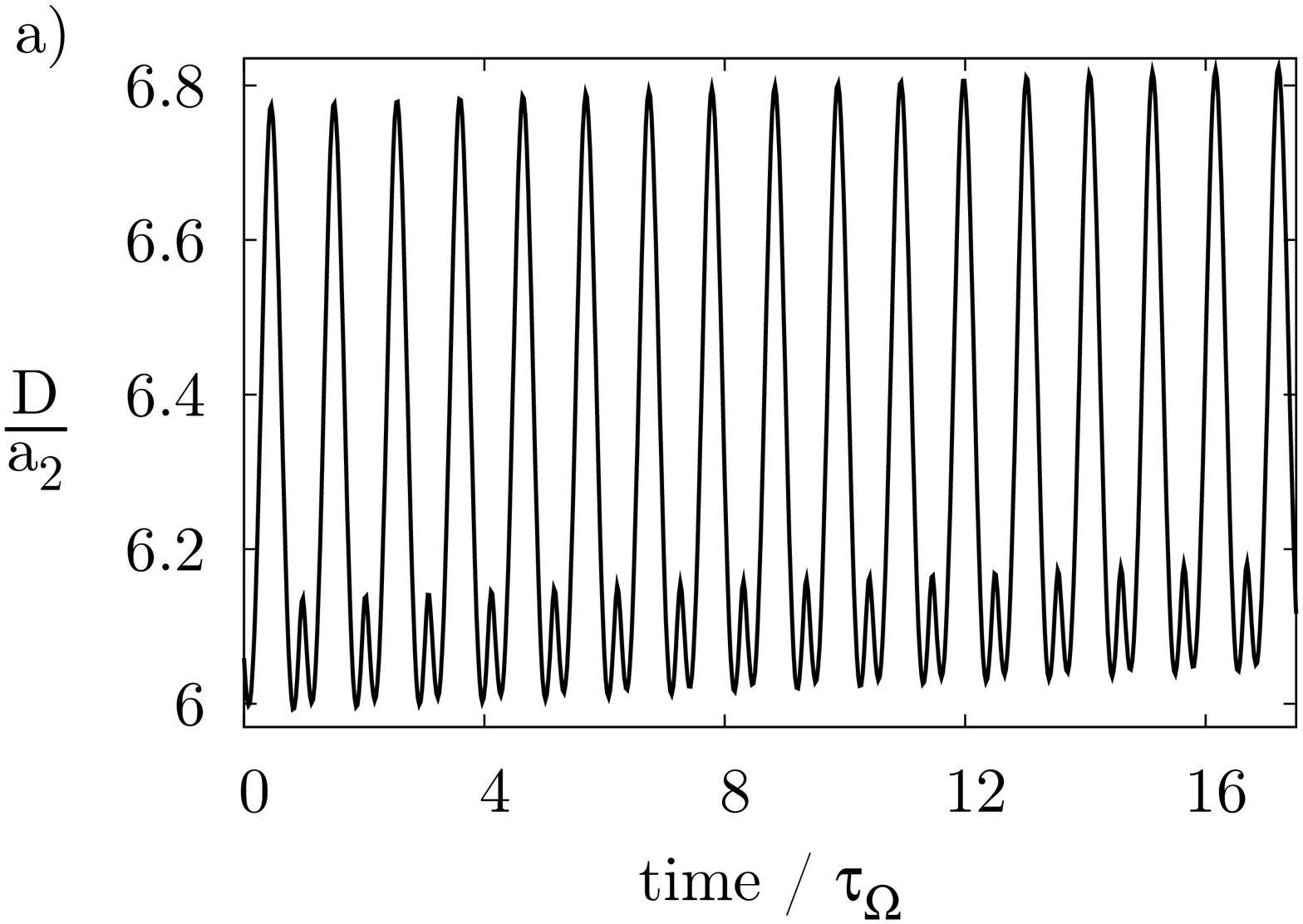}
     \hspace{2mm}
    \includegraphics[width=0.34\columnwidth, angle=-90]{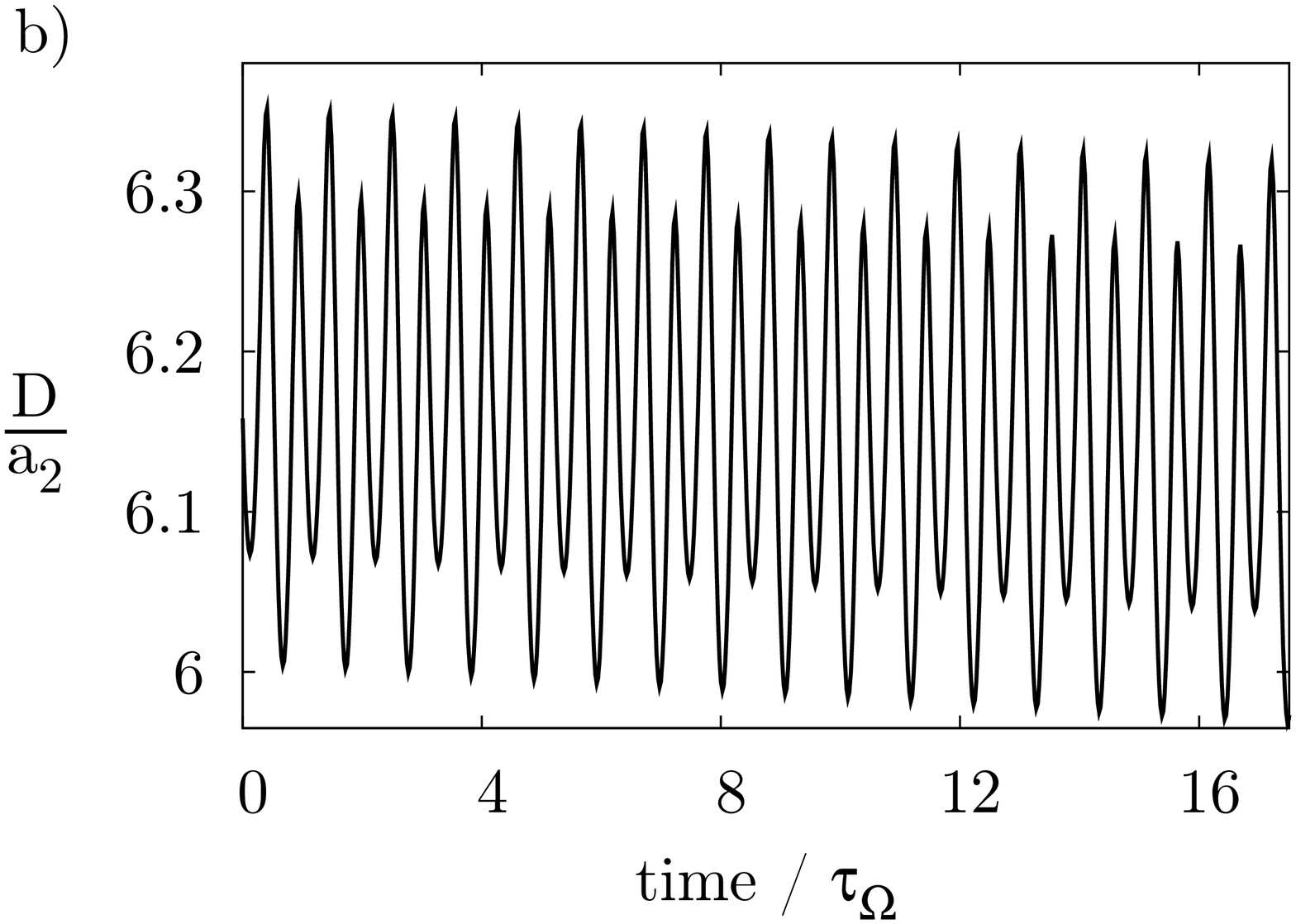}
  \end{center}
\vspace{-5mm}
  \caption{The distance $D(t)$ between
the centers of the  dumbbell is shown as a function
of the time for the same two parameter sets as
in figure \ref{ps_2_dumb}.  In part (a) the dumbbells repell each other. The parameters are
 $a_1 = 1.8$, $a_3 = 0.6$, $V_{21} = 1.2$ and $V_{43} = 1.0$. In part (b) the dumbbells attract each other for the parameters $a_1 = 1.2$, $a_3 = 1.1$, $V_{21} = 3.2$ and $V_{43} = 0.8$.
}
\label{distance_short_repulsion_attraction}
\end{figure}

\begin{figure}[ht]
\vspace{-3mm}
  \begin{center}
  \includegraphics[width=0.45\columnwidth, angle=-90]{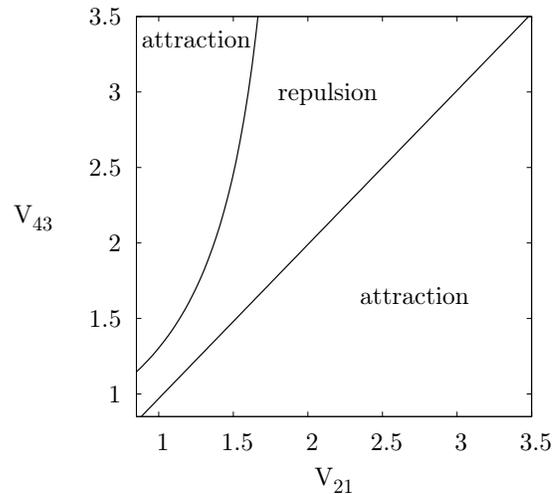}
  \end{center}
\vspace{-5mm}
  \caption{Ranges of dumbbell attraction and repulsion are shown
as a function of the coupling constants $V_{21}$ and $V_{43}$
for slightly differing dumbbell asymmetries: $a_1 = 1.2$, $a_3 = 1.1$.}
\label{phase_diagram_m1_m2}
\end{figure}

\begin{figure}[ht]
\vspace{-3mm}
  \begin{center}
    \includegraphics[width=0.45\columnwidth, angle=-90]{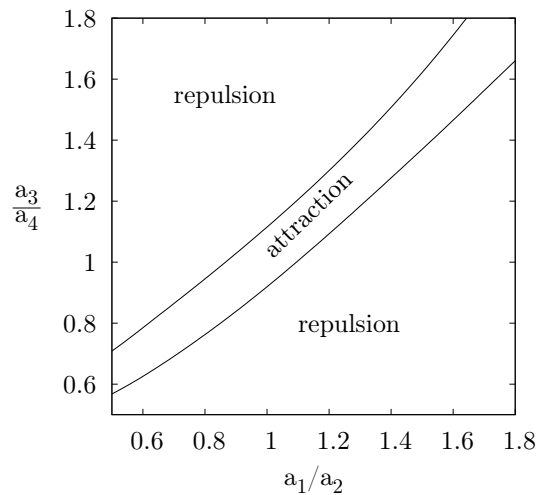}
  \end{center}
\vspace{-5mm}
  \caption{Ranges of dumbbell attraction and repulsion
are shown as functions of the ratios of the bead radii $a_1/a_2$ and $a_3/a_4$ for unequal coupling parameters, $V_{21} = 1.2$ and $V_{43} = 1.0$.}
\label{phase_diagram_a1_a3}
\end{figure}

\begin{figure}[ht]
\vspace{-3mm}
  \begin{center}
    \includegraphics[width=0.4\columnwidth, angle=-90]{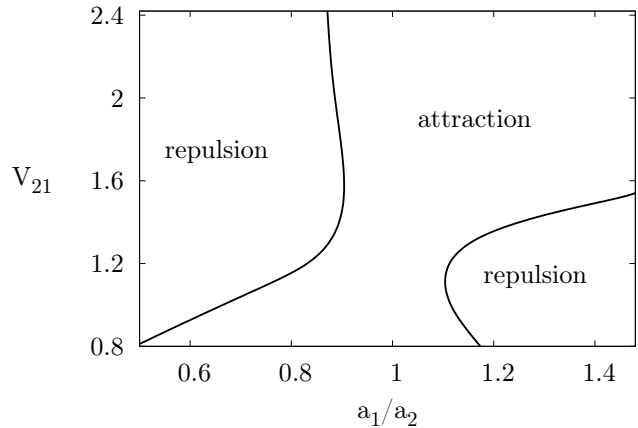}
  \end{center}
\vspace{-5mm}
  \caption{Ranges of dumbbell attraction and repulsion
are shown as a function of the coupling strength $V_{21}$
and the bead radius $a_1$. The other parameters are $V_{43} = 1.0$ and $a_3=1.1$.
}
\label{phase_diagram_a1_m1}
\end{figure}

As shown above the mean distance between two unequally shaped dumbbells may
increase, as sketched in
figure \ref{2_dumb_sketch1}(b), or decreases
as a function of time.
In figure \ref{distance_short_repulsion_attraction} we have plotted $D(t)$
for the same parameter sets as in figure \ref{ps_2_dumb}, but
for a longer time window. One can easily see
that in the mean the dumbbells repel each
other in part a) and attract each other in part b).
How the hydrodynamically induced repulsive or attractive dumbbell motion
depends on the parameters, describing the system's asymmetry, is
illustrated in terms of
diagrams for representative sets of parameters
in figures \ref{phase_diagram_m1_m2}-\ref{phase_diagram_a1_m1}.

In figure \ref{phase_diagram_m1_m2} areas of dumbbell attraction and
repulsion are shown as a function of the two coupling
parameters $V_{21}$ and $V_{43}$
for the fixed parameters,
 $a_1 = 1.2$ and  $a_3 = 1.1$, corresponding to
unequal dumbbell shapes.
For combinations of the two couplings along the
solid lines in figure \ref{phase_diagram_m1_m2}, the bead asymmetries of
the dumbbells are compensated by the difference between
$V_{21}$ and $V_{43}$, so that the
distance $D(t)$ is constant in the mean.
If one coupling parameter is much larger than the other one ($V_{21} \gg V_{43}$ or $V_{43} \gg V_{21}$)
the dumbbells attract each other. Between those two regimes there is a range of parameters for which the dumbbells
repel each other. For increasing coupling constants the repulsive area widens.

In figure \ref{phase_diagram_a1_a3} the regions
of dumbbell attraction and repulsion are shown as functions of
the two bead radii $a_1$ and $a_3$ for fixed
coupling constants, $V_{21} = 1.2$ and $V_{43} = 1.0$. For these parameters
there is only a narrow domain, in which the dumbbells attract each other.
In this area the dumbbell asymmetries
differ only about $20-30\%$. The phase diagram in figure \ref{phase_diagram_a1_a3} also shows
that for identical asymmetries of the two dumbbells,  $a_1=a_3$,
one may have either dumbbell repulsion, as in the case
 $a_1=a_3=0.6$, or attraction, as in the case $a_1=a_3=1.8$.

In figure \ref{phase_diagram_a1_m1} the coupling $V_{43} = 1.0$ and the asymmetry $a_3/a_4=1.1$ of one of the dumbbells were fixed, whereas the asymmetry $a_1/a_2$ as well as the coupling parameter $V_{21}$ is varied.
For a symmetric or slightly asymmetric dumbbell, i.e. $a_1/a_2 \approx 1$, there is always attraction, but for stronger asymmetries
there are also regions, where the two dumbbells repel each other.
Even for equal couplings ($V_{21} = V_{43}$) one can find areas of dumbbell attraction as well as regions in which the dumbbells repel each other.

\begin{figure}[ht]
\vspace{-3mm}
  \begin{center}
  \includegraphics[width=0.4\columnwidth, angle=-90]{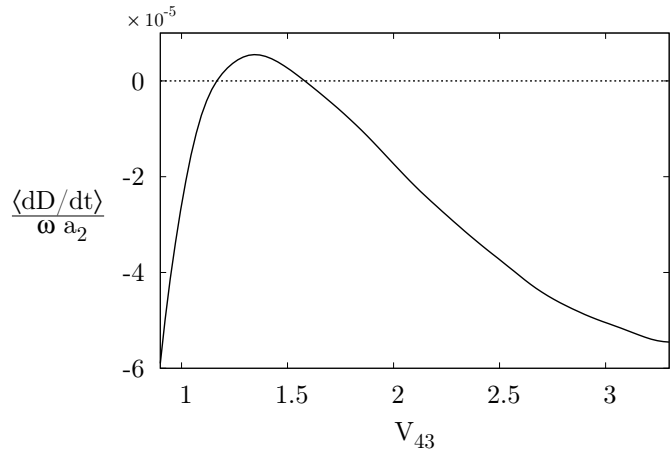}
  \end{center}
\vspace{-5mm}
  \caption{The mean relative velocity between the dumbbells is shown as a function of the coupling strength $V_{43}$ for the same parameter set as in figure \ref{phase_diagram_m1_m2} and the fixed value $V_{21} = 1.2$.}
\label{slice_through_pd_m1_m2}
\end{figure}

Away from the solid lines of vanishing attraction and repulsion in figure \ref{phase_diagram_m1_m2}, \ref{phase_diagram_a1_a3} or \ref{phase_diagram_a1_m1} the modulus of the relative velocity between the dumbbells increases with the distance from these lines. This is demonstrated in figure \ref{slice_through_pd_m1_m2}, where the mean relative velocity between the dumbbells is shown as a function of the coupling $V_{43}$. In this figure the same parameter set as
in figure \ref{phase_diagram_m1_m2} was used and $V_{21} = 1.2$ was fixed. In this case the transitions from attraction to repulsion and back occur at the values $V_{43} \approx 1.17$ and $V_{43} \approx 1.59$ and there is only a small range of values of $V_{43}$ for which the dumbbells repel each other.

Is there a simple mechanism by which the attractive or repulsive behavior
can be explained?
As shown before, the broken time reversal symmetry of $D(t)$ is
necessary for the non-reciprocal motion of the dumbbells to occur.
Furthermore for some parameter sets the phase shift $\Delta \phi$ between the dumbbells
as well as its modulation amplitude has a minimum in the repulsive
regime. However, this depends very much on the parameters being used.
For other parameter sets $\Delta \phi$ takes its maximum value
in the range in which the dumbbells repel each other.
For a third class of
parameter sets there is no obvious
correlation between the phase shift and the fact that the dumbbells behave in an attractive or a repulsive way.
All in all the mechanism leading to dumbbell attraction or repulsion
is a complex interplay between
the applied torques resulting from the coupling parameters and the drag forces,
which depend on the bead radii and thus on the asymmetries of the dumbbells.

\section{Conclusions and discussion}\label{sec: conclusions}

Hydrodynamic interactions between two rotating asymmetric dumbbells in
a fluid at a low Reynolds number were investigated. We found
either a temporally averaged
attraction or repulsion between the dumbbells if they
have different shapes and/or
their couplings to the driving field differ.
Differences in shape and/or
coupling break the time reversal symmetry, and are hence
 a pre-condition for attraction or repulsion. 
No generic rule could be identified of whether
a specific form of symmetry breaking leads to dumbbell attraction
or repulsion. 
We presented phase diagrams separating parameter regions of attraction from regions of repulsion.

We suggest experiments where anisotropic birefringent particles
are
rotated by circularly polarized light, similar to recent
experiments \cite{Chaikin:2002.1,Heckenberg:2004.1}, in order to
 explore the hydrodynamic attraction and repulsion
between
rotating small particles.
In such experiments the torque on an anisotropic particle
may easily be tuned by varying the laser power. Different torques
may also be applied to the particles by using different
shapes or materials for the particles. A spatially
varying power of the circularly polarized light can also cause different
torques on neighboring particles.

Dumbbells or anisotropic particles
can also be constructed from super-paramagnetic particles which
can be rotated using a magnetic field. In this case
the magnetic dipole-dipole interactions have to be taken into account.
Magic angle spinning may suppress the magnetic dipole or multipole interactions \cite{TFischer:2009.1} such that the hydrodynamic interactions dominate.

If the hydrodynamic interactions of an assembly of asymmetric objects
are investigated, collective dynamics such as chaotic motion can be expected. 
Our findings might find applications in an efficient way to micromix a fluid or to separate particles on the small scale.
It might also be useful for studies on the collective assembly of micro-swimmers.

\ack
We would like to thank J. Bammert for instructive discussions.
This work has been supported by the German science foundation through
the priority program on micro- and nanofluidics SPP 1164 and the
research group FOR 608.

\end{document}